\documentclass{svjour3}                     
\smartqed  
\usepackage{graphicx}
\newcommand{\be}{\begin{equation}}
\newcommand{\ee}{\end{equation}}

\newcommand{\es}{erg~s$^{-1}$}

\newcommand{\gtsima}{$\; \buildrel > \over \sim \;$}
\newcommand{\ltsima}{$\; \buildrel < \over \sim \;$}
\newcommand{\prosima}{$\; \buildrel \propto \over \sim \;$}
\newcommand{\gsim}{\lower.5ex\hbox{\gtsima}}
\newcommand{\lsim}{\lower.5ex\hbox{\ltsima}}
\newcommand{\simgt}{\lower.5ex\hbox{\gtsima}}
\newcommand{\simlt}{\lower.5ex\hbox{\ltsima}}
\newcommand{\simpr}{\lower.5ex\hbox{\prosima}}

\newcommand{\lx}{$L_{\rm X}$}
\newcommand{\lr}{$L_{\rm r}$}

\newcommand{\maxi}{MAXI~J1836--194}
\newcommand{\etal}{et~al.}
\begin{document}

\title{An overview of jets and outflows in stellar mass black holes
}
\subtitle{}


\author{Rob Fender         \and
        Elena Gallo 
}


\institute{R. P. Fender\at
              Astrophysics, Department of Physics, University of Oxford, Keble Road, OX1 3RH, Oxford, UK\\
              \email{rob.fender@astro.ox.ac.uk}           
           \and
           E. Gallo\at
              Department of Astronomy, University of Michigan, 500 Church St., Ann Arbor, MI 48109, USA
}

\date{Received: date / Accepted: date}

\maketitle

\begin{abstract}
In this book chapter, we will briefly review the current empirical understanding of the relation between 
accretion state and and outflows in accreting stellar mass black holes. The focus will be on the
empirical connections between X-ray states and relativistic (`radio') jets, although we are now also
able to draw accretion disc winds into the picture in a systematic way. We will furthermore consider
the latest attempts to measure/order jet power, and to compare it to other (potentially) measurable
quantities, most importantly black hole spin. 
\end{abstract}

\section{Introduction}
\label{intro}

Jets, collimated relativistic outflows carrying large amounts of
energy away from the deepest parts of the gravitational potential
well, are a relative latecomer to the overall picture of accretion
around stellar mass black holes. Much of the basics of accretion
theory, namely how the matter gets in while angular momentum and
radiation get out, was developed in between the 1960s and 1980s, with
roots in much earlier works, and is still applicable today
(and much of it will have been covered in other chapters in this book). 
In contrast, it took until the 2000s for it be accepted that jets are a
key part of the accretion process in stellar mass accretors, both in
terms of their uniquity and importance in having very large kinetic
powers. Furthermore, despite knowledge of their existence in active
galactic nuclei for a century, the basics of how relativistic jets are
launched are still a long way behind our understanding of the
accretion process.

\begin{figure*}
\hspace{0.2in}
  \includegraphics[width=0.95\textwidth]{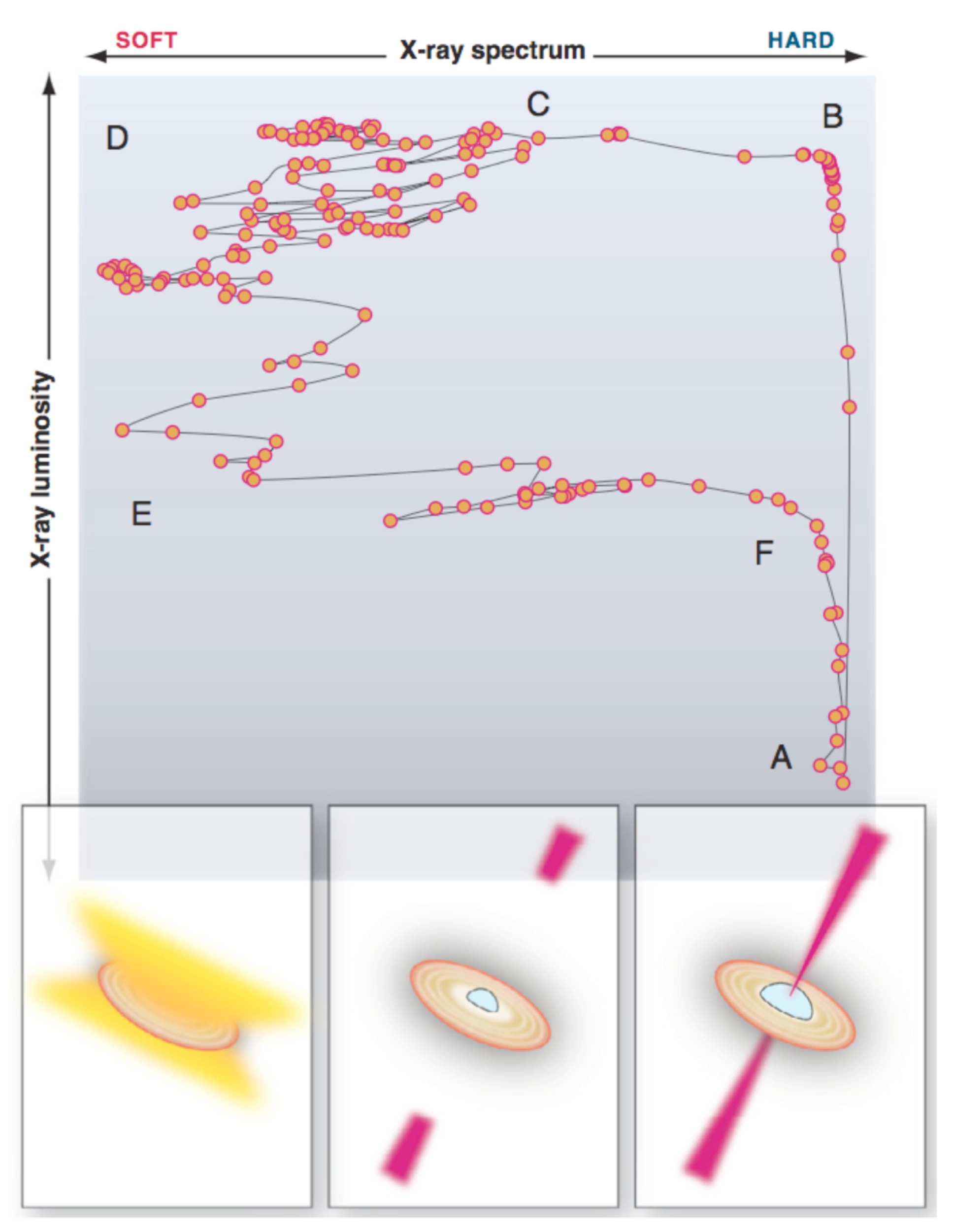}
\caption{A qualitative summary of the relation between accretion states and
outflows in black hole X-ray binaries. The upper panel represents the relation
between X-ray spectral hardness and luminosity, and the path A-B-C-D-E-F 
represents the path taken by the recurrently-outbursting binary GX 339-4 over
the course of about one year. Hard states (to the right) are well-studied and
show quasi-steady radio jets. Soft states (to the left) appear to have no
radio jets but strong accretion disc winds. Transitions between the states, at least at high
luminosity (e.g. B-C-D), appear to be associated with discrete jet ejection events.
From Fender \& Belloni (2012), where a concise summary of the cycle can be found.
}
\label{science}      
\end{figure*}

Nevertheless, once we discovered that jets in X-ray binaries follow
certain patterns, revealed mainly (but not exclusively) by coordinated
observing campaigns in the radio and X-ray bands, we realised that
they offered perhaps our best opportunity to understand the connection
between accretion and jet formation in relativistic objects.

In this paper we will describe the current state of play in our
empirical understanding of the relation between accretion, jets and
winds in stellar mass black holes (and neutron stars). We will focus
on thoroughly summarising the observational evidence, and highlighting
recent progress.

For a lengthier introduction to the basic observables of jets 
and what can be understood from them, 
the reader is directed towards
Fender (2006) and references therein. For a very quick summary of the
state of the field, see Fender \& Belloni (2012). Other relevant
historical reviews include Hjellming \& Han (1995), Mirabel \& Rodr{\'{\i}}guez (1999),
Mart{\'{\i}}\ (2005).

\section{The empirical picture}
\label{sec:1}

In this section we shall summarise the empirical evidence for a
connection between accretion and outflow in X-ray binaries. We shall
start with the black hole X-ray binaries, for which the evidence is
strongest. Fig \ref{science} summarises both the current state of our knowledge of
the connection between accretion and outflows in black hole X-ray binaries,
and how we most commonly choose to represent it. This representation of the 
relation between jets and X-ray state originated with Fender, Belloni \& Gallo (2004), and
has since been greatly augmented by the work of Ponti et al. (2012) who found
a clear relation between accretion disc winds and the soft X-ray state.

\begin{figure*}
  \includegraphics[width=1.0\textwidth]{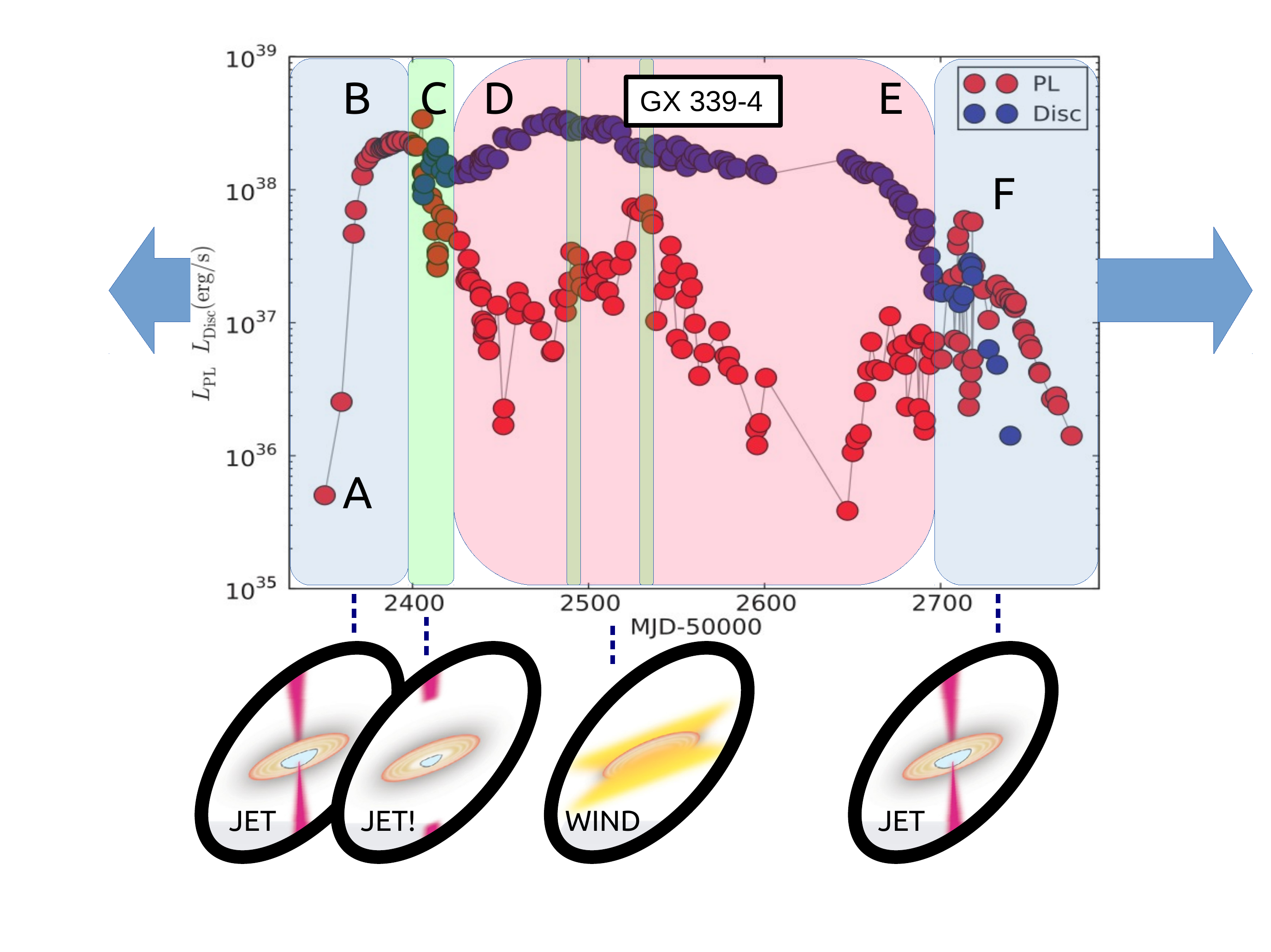}
\caption{An outburst of GX 339-4, this time shown as temporal evolution.
The red and blue curves correspond to power-law and disc components of
the X-ray spectrum: blue-dominated states are `soft' and red are `hard',
in the context of Fig 1. The figure illustrates that the majority of the
time at the highest luminosities is spent in the soft state where
there appears to be non core radio jet, but instead a strong wind. 
The letters correspond approximately to those in Fig 1.
Partially adapted from Dunn et al. (2010) and Fender \& Belloni (2012).
}
\label{lightcurve}      
\end{figure*}

We are very fortunate that, on timescales fairly well tuned to PhD
durations or grant cycles, we can observe a single stellar mass black
hole in a binary system undergo accretion in a small number of
different modes over a large range of luminosities. The cycle illustrated
in Fig \ref{science} takes place over about one year, as is illustrated
in Fig \ref{lightcurve}. Noteworthy points to take from this figure are
that the quasi-steady jet is present most of the time, although mainly
at low luminosities, that the transient/flaring jet phase constitutes
a very small fraction of the outburst, and that for most of the time
at the highest luminosities there appears to be no core jet but a strong
disc wind.

The following subsections refer to the (quasi-cyclic) 
phases labelled A through F in Figs 1 and 2.

\subsection{A to B (and F to A): quiescence to bright hard states}

Outbursts of black hole X-ray binaries (BHXRBs) seem to be caused by a
switch in the viscosity of the accretion disc, associated with the
ionisation of hydrogen. The most recent test of this theory,
originally developed for white dwarf accretion in cataclysmic
variables (Meyer \& Meyer-Hofmeister 1981), continues to support the scenario (Coriat, Fender \& Dubus
2012). Once the viscosity increases, the accretion rate at the inner
edge of the accretion disc rises dramatically. Since this is where
most of the accretion luminosity is produced, the source gets brighter
on a short timescale (days).

In our experience to date, most BHXRBs spend most of their time
outside of outbursts at `quiescent' levels of accretion with
corresponding luminosities of $\sim 10^{30-32}$ erg s$^{-1}$. This
population is not very well sampled, however, and some sources are
stable at higher luminosities (e.g. V404 Cyg which is `quiescent' at
$\sim 10^{34}$ erg s$^{-1}$) and selection effects suggests that
there's likely to be a tail to lower luminosities. Once the outburst
begins, however, the sources can rise to luminosities $\geq 10^{38}$
erg s$^{-1}$ very rapidly, an extremely dramatic change. For most of
this range in luminosity the X-ray spectrum 
remains in a `hard' spectral state (however, it is likely that most of the action is
going on in the ultraviolet, which is hard/impossible to observe in
for most X-ray binaries, due to large dust extinction in the galactic
plane). This state is characterised by a
broad-band X-ray spectrum with photon index $\Gamma \sim 1.6-2.1$ (with the spectrum hardening as the luminosity increases; e.g. Plotkin et al. 2014) often
observed to show a cutoff between $\sim$50-130 keV (with the temperature decreasing the luminosity increases; e.g., Motta et al 2009; Joinet et al. 2008), and generally ascribed to
thermal Comptonisation in a hot plasma or `corona' above/around/near
the accretion disc. There is usually strong X-ray variability (up to
40\% r.m.s. in the Fourier frequency range 0.01 -- 100 Hz) associated
with this state.

In this hard state, BHXRBs essentially always show relatively weak but
steady radio emission, with a flat ($\alpha \sim 0$, where flux
density $S_{\nu} \propto \nu^{\alpha}$) spectrum and low levels of
polarisation (Fender 2001; but see Brocksopp et al. 2013 and Russell \& Shahbaz 2014). 
The relative steadiness, flat spectrum and lack of
polarisation suggest that this emission originates in a more or less
continuously replenished, partially self-absorbed outflow, such as
those originally conceived of to explain the flat spectrum cores of
some quasars. The radio-X-ray correlation has been studied extensively,
and is discussed in more detail in the following subsection.

\subsubsection{Luminosity correlations}
\label{scaling}

\begin{figure*}
 \includegraphics[width=1.\textwidth]{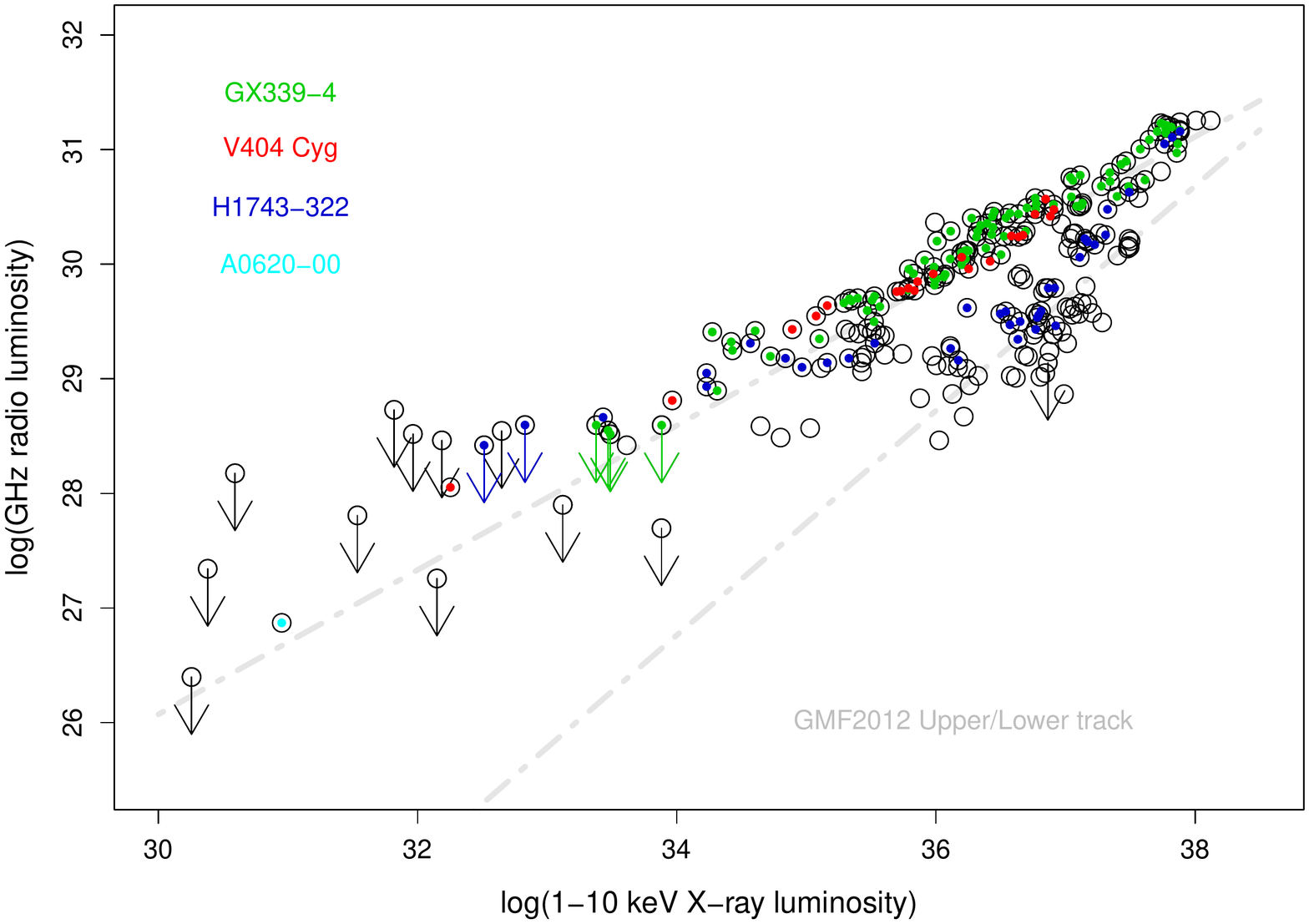}
\caption{The radio/X-ray luminosity plane of black hole X-ray binaries
  (c.a. 2014). Shown are quasi-simultaneous radio and X-ray luminosity
  measurements (in log CGS units) for 24 black hole X-ray binaries,
  ranging from quiescence (A0620$-$00; Gallo \etal\ 2006) up to bright
  hard states. The dashed grey lines indicate the best-fit relations
  to the upper and lower tracks as identified by Gallo \etal\ (2012)
  (\lr$\propto$\lx$^{0.63}$ and \lr$\propto$\lx$^{0.98}$,
  respectively). We also highlight the two sources for which a tight
  non-linear correlation of the form \lx$\propto$\lr$^{0.6-0.7}$ has
  been reported over a wide dynamic range (i.e. GX339--4; Corbel
  \etal\ 2003, and V404 Cyg; Gallo Fender Pooley 2003; Corbel,
  K\"ording \& Kaaret 2008), plus the enigmatic source H1743--322,
  which appeared to ``jump" from the lower to the upper track as it
  faded into quiescence following its 2008 outburst (Jonker
  \etal\ 2010; Miller-Jones \etal\ 2012). }
\label{fig:plane}      
\end{figure*}

Quasi-simultaneous radio and X-ray monitoring has become the standard
tool of investigation for the so called ``jet-accretion coupling" in
hard state BHXRBs. A strong and repeating correlation
 has been established between the radio and X-ray luminosity for two systems, with \lx~ being proportional to \lr$^{0.6-0.7}$ 
(GX339$-$4: Corbel \etal\ 2003, 2013 and V404 Cyg: Gallo, Fender \& Pooley 2003; Gallo, Fender \& Hynes 2005, 
Corbel, K\"ording \& Kaaret 2008). However, the universality of this relation 10 years after
its discovery is far from obvious.  Fig \ref{fig:plane} summarizes
the current state of the problem by assembling what is likely the
most complete data collection as of today (data from Gallo, Miller \& Fender
2012 plus Corbel \etal\ 2013, and references therein). Two main
differences stand out with respect to the simple picture we used to
draw a decade ago: firstly, there appears to be {\it two luminosity tracks}
(see Gallo \etal\ 2012 for more details); secondly, the behavior of
the BHXRB H1743--22 (in blue) as observed during the decline of its
its 2008 outburst (Jonker \etal\ 2010; Miller-Jones \etal\ 2012; Miller \etal\ 2012) is in stark contrast with that
of GX339--4 and V404 Cyg (green and red, respectively): H1743--22
starts off as under-luminous in the radio band during the initial
outburst decay phase ($10^{36}\simlt $\lx$\simlt 10^{38}$\es),
proceeds to make a nearly horizontal excursion toward lower X-ray
luminosities (between $10^{36}\simlt$\lx$<\simlt10^{36}$\es), and
finally reaches a comparable radio luminosity level (for the same \lx)
as GX339--4 and V404 Cyg (this happens below $\simeq 10^{34}$ \es).
Whatever drives the relative radio quietness/loudness of BHXRBs in
hard and quiescent states, thus, cannot be linked to any physical property
that remains constant over typical outburst time-scales (e.g., black
hole spin, orbital inclination; see Soleri \& Fender 2011).  

Coriat \etal\ (2011) ascribe the erratic behavior of H1743$-$22, and by analogy
that of other radio-quiet systems, to the onset of a  
radiatively efficient inflow  (see also K\"ording, Fender \& Migliari 2006).
Along the same lines, a more theoretical basis for the existence of a radio-quiet (or rather,
X-ray-bright) track has been recently proposed by Meyer-Hofmeister \& Meyer (2014), who argue that thermal photons from a weak,
cool, inner disc (sustained by re-condensation of optically thin gas
within the inner regions; Meyer \etal\ 2007; Liu \etal\ 2007) could be
responsible for enhancing the seed photon field available for
Comptonisation, and hence the hard X-ray flux.  From a theoretical
standpoint (as well as observational, see e.g. Miller \etal\ 2006; Reis Fabian \& Miller 2010; Reynolds \& Miller 2013
and references therein), the inner cold disc would cease to exist at low
accretion rate. Thus, {\it{the radio-quiet track is not expected to extend down to the quiescent regime.}} Though this exact behavior
is just what was observed in the case H17143$-$22, whether the inner
disc re-condensation scenario will stand the test of time (in terms of
reproducing the X-ray and radio behavior of BHXRBs) depends critically
on our ability to probe the truly quiescent domain (see Calvelo \etal\ 2010 and Miller-Jones
\etal\ 2011 for a detailed discussion on how distance limitations are
likely to hamper a systematic investigation in the radio band). To date, A0620$-$00 is the only truly quiescent system with a robust radio detection, and its radio luminosity seems to lie on the extrapolation of the GX339$-$4 and V404 Cyg radio/X-ray correlations down to Eddington-scaled X-ray luminosities as low as $\sim 10^{-9}$ (Gallo \etal\ 2006). However, with a sample of one, and considering selection effects, this is a long way from conclusive.

At the same time, a systematic study on the radio properties (most notably spectral index and polarization degree) of BHXRBs while on the radio quiet/X-ray bright track could provide complementary information on the physics that drives this apparent divide (there are hints that the ``radio quiet" objects have slightly more optically thin radio spectra, Brocksopp et al. 2013). 
A good candidate for the origin of the flat spectrum, left unresolved in Blandford \& K\"onigl (1979), is internal shocks
driven at the frequencies of the X-ray variability (Malzac 2013).
\\

Determining the compact jet contribution at mm, infrared (IR) and optical frequencies is admittedly challenging; much progress has been made in this direction over the last 10 years. A growing number of works (based both on spectral energy distribution analysis as well as variability studies) support the claim that, while in the hard state, the compact jet extends all the way up to IR and often optical frequencies (Fender 2001; Chaty \etal\ 2003; Bradley \etal\ 2007; Gallo \etal\ 2007; Russell \etal\ 2010, 2011b; Brocksopp \etal\ 2010; Malzac \etal\ 2004; Hynes \etal\ 2004, 2006, 2009; Markoff \& Nowak 2007; Casella \etal\ 2010). Most relevant to this section, Russell \etal\ (2006) collected nearly-simultaneous IR/optical and X-ray observations of 33 X-ray binaries (black holes and neutron stars) to estimate the relative contributions of various IR/optical emission processes as a function of X-ray luminosity. They found evidence for a positive correlation between the IR/optical and X-ray luminosity, of the form $L_{\rm opt-IR}\propto$\lr$^{0.6}$,  extending all the way from the peak of the hard state down to the loses quiescent X-ray luminosities. Notwithstanding the large scatter, no strong outliers, exceptions, sub-clusters or bifurcation have been reported so far for this relation. 

\subsection{B-C-D: Transition from hard to soft}
Once the rising source passes an X-ray luminosity of $\sim 10^{37}$
erg s$^{-1}$ (albeit with substantial variations between sources
and even between different outbursts of the same source; see, e.g., Dunn et al. 2010 ), at which level it will still be in the hard state, it is
very likely to make a subsequent transition to a softer spectral state. 
The transition from the hard state to the soft state can occur with very little change (of order unity) in the broadband luminosity of the system (in spite of rather dramatic changes in the spectral shape), and has a number of very interesting characteristics. For a `typical' BHXRB the spectral transition
from hard to soft states can take days or weeks, but the evolution can be much more dramatic in terms of the X-ray power spectra.
While the hard and soft states appear to be dominated by strong ($\geq 30$\%) and very weak ($\leq 5$\%) broad band noise respectively, some phases during the transition can be completely dominated by strong peaked and QPO-like components. Casella et al. (2005) classified these QPOs into three types: A, B \& C, with strong type-Bs being the most indicative of the transition.

The hard to soft transition is also where the brighest radio flares occur, sometimes singly and sometimes in sequences.
These radio flares can peak, in the 1--5 GHz band, at many times the preceding hard state flux density, and (mostly) evolve
in a `standard' (e.g. van der Laan 1966) way from optically thick to thin as they (presumably) expand. Such flares are
occasionally spatially resolved by radio telescopes into individual radio knots which move away from the radio core (presumably
the site of the black hole) at (at least) mildly relativistic speeds (e.g. Mirabel \& Rodr{\'{\i}}guez 1994; Hjellming \& Rupen 1995; Fender et al. 1999; MIller-Jones et al. 2012).
It appears, although it is not at all proven, that these jets may move faster than those in the hard state, and so what we may be
seeing is at least partially associated with internal shocks in a flow of variable (increasing) speed (Fender et al. 2004 and references therein; Fender et al. 2009). However, it is certainly not clear that they are significantly more {\em powerful} than the preceding,
bright, hard state jets.

There have been a number of attempts to associate the `moment of jet launch' with an associated change or event in the accretion flow.
Fender, Homan \& Belloni (2009) and Miller-Jones et al. (2012) both attempted to see if the momenet of jet launch could be associated with
changes in the timing (X-ray variability) properties of a small number of black hole binaries, in particular with the occurrence of the type-B QPOs (see also discussion in Soleri, Belloni \& Casella 2008). While they are certainly broadly coincident, a one-to-one relation could not be firmly established and whether or not there exists a key signature of the moment of launch remains unclear. Seemingly at odds with this statement, in the bright black hole binary GRS 1915+105, which undergoes long cycles of accretion state changes on timescales as short as minutes, it is clear that each cycle is associated with an ejection event, and possibly even with a particular moment in that cycle (e.g. Mirabel et al. 1998; Klein-Wolt et al. 2001). However, these state changes appear to be compressed versions of what happens on longer timescales in other systems, and the blurring effect caused by the size scale of the radio emission makes direct association with some change difficult. Success is more likely in the infrared band, where much effort was made for GRS 1915+105 (e.g. Fender et al. 1997; Eikenberry et al. 1998; Mirabel et al. 1998) but it is extremely hard to maintain minutes-resolution monitoring of a more normal black hole binary across its entire week-long transition phase. Finally, it should be noted that there may be no clear `moment of jet launch' at all: internal shock models are showing promise in reproducing some of the radio properties of X-ray binaries (e.g. Jamil \etal\ 2010; Malzac 2013) and it may be that a rapid, but not instantaneous, change in jet properties (such as injection cycle time or speed) are enough to reproduce what we see as bright flares in the radio band.

\subsection{D to E: The soft state}

In the soft state the X-ray spectrum is dominated by a blackbody-like
component which peaks around 1 keV, combined with a steeper and weaker
power law. The origin of the thermal component is likely to be an
optically thick accretion disc. The overall X-ray variability drops
dramatically in this state (often below 5\%). The core radio, mm and near-infrared emission drop dramatically in the
soft state (Fender et al. 1999; Russell et al. 2011a), the
simplest interpretation of which is that the core jet has switched off.

A major breakthrough in understanding states appears to have arrived recently
with Ponti et al. (2012) demonstrating that accretion disc winds,
revealed in X-ray spectra, appear to be uniquely observed in {\em edge-on, soft-state} BHXRBs
(caveat relatively poor sampling of late-time soft states, and some remaining small
uncertainty about how evolving ionisation could affect this conclusion).
Thus it seems that in moving from the hard to the soft state, we leave a regime
of strong, quasi-steady jets and little, if any accretion disc wind and enter
the converse regime, with strong winds and weak jets (see also Miller \etal\ 2006, 2008; Neilsen \& Lee 2009).
These two regimes are probably not simply a rebalancing of the same outflow
power, however, with the wind probably carrying less kinetic power but more mass
than the hard state jet (detailed calculations of kinetic energy, mass and
momentum flux in these two modes have yet to be carried out, however).

\subsection{X-ray states: other views and potential origins}

The X-ray states described above have been done so from a viewpoint which is 
very much biased towards the outflows. Other very useful reviews, usually more focussed
on the X-ray emission and what it tells us about the workings of the inner
accretion flow, can be found in Remillard \& McClintock (2006), Done, Gierlinski \& Kubota (2007), 
Belloni, Motta \& Munoz-Daris (2011). The general pattern of hysteresis can
also be viewed in other ways than the {\em hardness-intensity} diagram used in Fig 1. 
In particular Munoz-Darias, Motta \& Belloni (2011) and Plant et al. (2014) 
utilised {\em rms-intensity} and {\em reflection-intensity} diagrams respectively to show broadly
the same patterns but shedding new light on the detailed evolution of the accretion flow (see also Belloni \& Stella, this book).

It should be noted that while the root cause of the outburst may be agreed upon
as being the hydrogren ionisation instability, the origin
of the accretion state cycles and associated spectral hysteresis is not at all clear.
There are published models which may be testable in the future
(e.g. Petrucci et al. 2008; Begelman \& Armitage 2014 and references therein),
and very similar cycles of behaviour appear to be present in systems containing
both neutron stars and white dwarfs (K\"ording et al. 2008b; Munoz-Darias et al. in prep).

There are good reasons to believe that the patterns of behaviour observed in stellar-mass
black holes should be scalable to intermediate mass black holes and supermassive black
holes in active galactic nuclei, in fact there is strong observational support for this
view. Merloni, Heinz \& Di Matteo (2003) and Falcke K\"ording \& Markoff (2004) presented
the first evidence for relatively neat scalings of X-ray/radio luminosities and black hole mass,
across the whole black hole mass range. McHardy \etal\ (2006) and K\"ording \etal\ (2007)
made similar analyses for timing frequencies. K\"ording, Jester \& Fender (2006) made
an argument that the overall patterns of coupling between states and jets might be
the same in AGN, although this connection has really yet to be established convincingly (see also K\"ording, this book).

\section{Jet power}

Radiative and mechanical feedback from relativistic jets is thought to
play a key role in regulating the growth of galactic bulges and their
nuclear super-massive black holes. Empirical scaling relations between
the mass of super-massive black holes in nearby galaxies and large
scale properties of their host bulges, such as the well know
``black-hole-mass/stellar velocity dispersion" relation (see Kormendy
\& Ho 2013 for a recent review) are interpreted as strong
observational evidence for strong co-evolution. At the same time,
state-of-the-art cosmological simulations (e.g. Khandai \etal\ 2014)
that trace the assembly and merger history of galaxies and their nuclear
black holes rely on some form of black hole- (plus supernova) -driven
feedback in order to reproduce them.  Nevertheless, even the most
sophisticated models make use of over-simplified recipes for the
relative efficiencies of accretion vs. jet power, and {\it posit} that
the jet power, at least during the quasar phases, is set by the spin
of the black hole.  As we discuss in the following sections,
BHXBRs can actually help shed light on some of these processes.

\subsection{Jet power and black hole spin}

The holy grail for relativistic jet astrophysics is to establish the
jet formation mechanism. More specifically, we would like to be able
to identify observational signatures pointing unequivocally to models
that rely solely on differential rotation (Blandford \& Payne 1982)
vs. models that tap directly into the rotational energy of the black
hole (Blandford \& Znajek 1977; BZ). While super-massive black holes
in AGN offer the advantage of vast demographics, BHXRBs in soft states
provide us with a more favorable environment to systematically employ
X-ray spectral fitting techniques to measure the temperature, and thus
the extent, of the inner accretion disc (see Miller 2007 and Reynolds 2013 for 
recent reviews, as well as Reynolds, and McClintock, Narayan \& Steiner, this book). This translates into an estimate of the black hole
spin parameter $a$, which can then be compared against the jet power
$P_{\rm j}$ in order to (dis)prove a relation of the form $P_j\propto
a^2$, predicted by the BZ model (see 	
Tchekhovskoy \etal\ 2010 for a full relativistic traetment). 

How to measure $P_{\rm j}$?  When it comes to BHXRBs, persistent, hard
state, compact jets are a natural place to start.  Unfortunately, that
the face-value radio luminosity of compact jets is at best a poor
indicator of jet power is apparent from a number of considerations;
firstly, although compact jets seem to be persistently on during hard
states, their flux density varies by orders of magnitude, with some
non linear power of \lx~(see section \ref{scaling}). In addition, it should be kept in mind that
any inference of radio luminosity that is based on a single-frequency
flux density measurement relies on assuming a specific spectral shape
at lower frequencies (often taken as flat). Further, extrapolating the
integrated radio luminosity to total (i.e. radiative plus kinetic) jet
power relies on even more crucial assumptions: $i)$ on the location of
the optically-thin-to thick jet break (which is known to vary with
overall luminosity, yet not necessarily as basic scaling relations
would predict; see next section); $ii)$ on the location of the cooling
break (possibly constrained in one source, see again next section)
$iii)$ on the radiative efficiency of the synchrotron process.

Based on the above arguments, in attempt to test the presence of a
correlation between jet power and spin parameter in hard state BHXRBs,
Fender, Gallo \& Russell (2010) adopted a phenomenological approach
whereby the relative normalization of the (thought to be universal)
radio/X-ray and/or infrared/X-ray correlation were taken as a proxy
for total jet power. Perhaps not surprisingly, no evidence for a
positive correlation between these normalisations and reported spin
measurements emerged -- a result that could be interpreted as
due the large uncertainties in measuring jet power (and, to a second
extent, spin).  A different approach was taken by King \etal\ (2013a),
who found a marginally significant positive correlation between the
mass-scaled radio luminosity and spin parameter across a sample of 11
BHRBs and 37 Seyfert galaxies.

Ideally, one would like to calibrate the radio luminosity-to-total jet
power relation conversion via robust observational constraints on the
amount of work exerted by the jets on the surrounding inter-stellar
medium, and use jet-inflated radio lobes and cavities as effective jet
calorimeters, similarly to what has been done for super-massive black
holes in radio galaxies for decades. Unfortunately, the number of
jet-powered large-scale structures is very slim for BHXRBs, mainly as
a result of the jets propagating though under-dense environments
compared to AGN (see Heinz 2002). Albeit the small number statistics, these large scale structures point 
towards the jets carrying away a substantial fraction of the overall
accretion energy budget (Heinz \& Grimm 2005; Gallo \etal\ 2005;
Fender, Maccarone \& van Kesteren 2005; K\"ording, Jester \& Fender 2008; Heinz, Merloni \& Schwab 2007, King \etal\ 2103b). \\

More recently, much excitement has grown around the claim of a
positive correlation -- consistent with the BZ-predicted scaling  -- between spin parameter and jet power {\it for
  transient jets} (Narayan \& McClintock 2012; Steiner, McClintock \& Narayan 2013). In these works, the mass-scaled peak radio luminosity of the giant
radio flare that is often observed during hard-to-soft state
transitions is taken as proxy for jet power (while the spin
measurements rely on fitting the broadband X-ray spectrum in pure
thermal states).  In a subsequent work, Russell, Gallo \& Fender (2013)
argued against there being a significant correlation. While the
controversy is still open, much of the uncertainty depends on the
relative scarcity of observations of Eddington-limited BHXRBs, which
are supposed to act as ``standard candles" (cf. Steiner
\etal\ 2013). The debate will likely settle over the next few years,
as more and better data become available; in the meantime, the number
of possible exceptions to the $P_{\rm j}\propto a^2$ scaling relation for transient jets may be growing; the extragalactic BHXRB in M31 (Middleton, Miller-Jones \& Fender 2014), and the Galactic system 4U 1630--472
(King \etal\ 2014) seem to display too high/low (respectively) flare
radio luminosity for their reported spin values (unless exceptionally 
high values of jet boosting/de-boosting factor are at play). For the ``microquasar" in M31 (Middleton \etal\ 2013), however,
the significant uncertainty on the system inclination also allows for a
reconciliation with the best-fit scaling relation reported by Steiner
\etal\ (2013). For 4U 1630--472 , the high spin parameter reported by
King \etal\ (2014) relies on fitting the X-ray reflection spectrum rather than
thermal continuum spectrum. While the number of discrepancies between the
two spin-measuring methods is reported to be diminishing (see discussions in Miller \etal\ 2009; Fabian \etal\ 2012), the consensus is that any convincing relation between jet power and spin
parameter ought to be verified by both (or, at least, that either
spin-fitting method leading to the existence of a positive relation
with jet power does not constitute a legitimate argument for
invalidating the other).  \\

An entirely new approach to measuring black hole spin has been recently devised by Motta \etal\ (2014a,b); rather than fitting X-ray energy spectra, this technique relies on the relativistic precession model quasi-period oscillations (QPOs) in the power density spectra of BHXRBs (Stella \& Vietri 1999). 
This method has been successfully applied to the BHXRB GRO J1655-40 (Motta \etal\ 2014a) and XTE J1550-564 (Motta \etal\ 2014b). In the former case, the recovered black hole mass value matches the known system's mass function, while the spin value is not consistent with X-ray spectroscopy. In the latter, the spin value estimated from the relativistic precession model is  consistent with X-ray spectroscopy (while the known black hole mass was used to solve for spin). Going forward, this could prove an exceptionally powerful method to infer black hole spin and mass values, in that it is not affected by many of the complexities of radiative models.

\subsection{Jet power and spectral breaks}
 
 As briefly discussed in the previous section, the location where the optically thick (partially self-absorbed) synchrotron spectrum breaks and becomes optically thin also determines the peak of the jet flux density. Standard synchrotron theory predicts this jet break
 frequency ($\nu_{b}$) to scale with the mass accretion rate and black
 hole mass (Heinz \& Sunyaev 2003; Markoff \etal\ 2001, 2003,
 2005). For BHXRBs, assuming that the magnetic field and jet
 acceleration region size do not vary dramatically, a positive
 relation should exist between $\nu_b$ and mass accretion rate.  In
 order to test for such a correlation (in the form of $\nu_b
 \propto~$\lx$^{1/3}$) Russell \etal\ (2013a) performed a comprehensive literature search for nearly simultaneous multi-wavelength (including
 radio, IR and optical) data of BHXRBs in quiescent and hard
 states. This work represents the most accurate and complete study on
 BHXBR jet spectral breaks to date, with 12 BHRXBs having constraints
 on $\nu_b$.  No global relation was found with \lx, from
 $10^{-8}$ up to the Eddington limit.

\begin{figure*}
 \includegraphics[width=1.\textwidth]{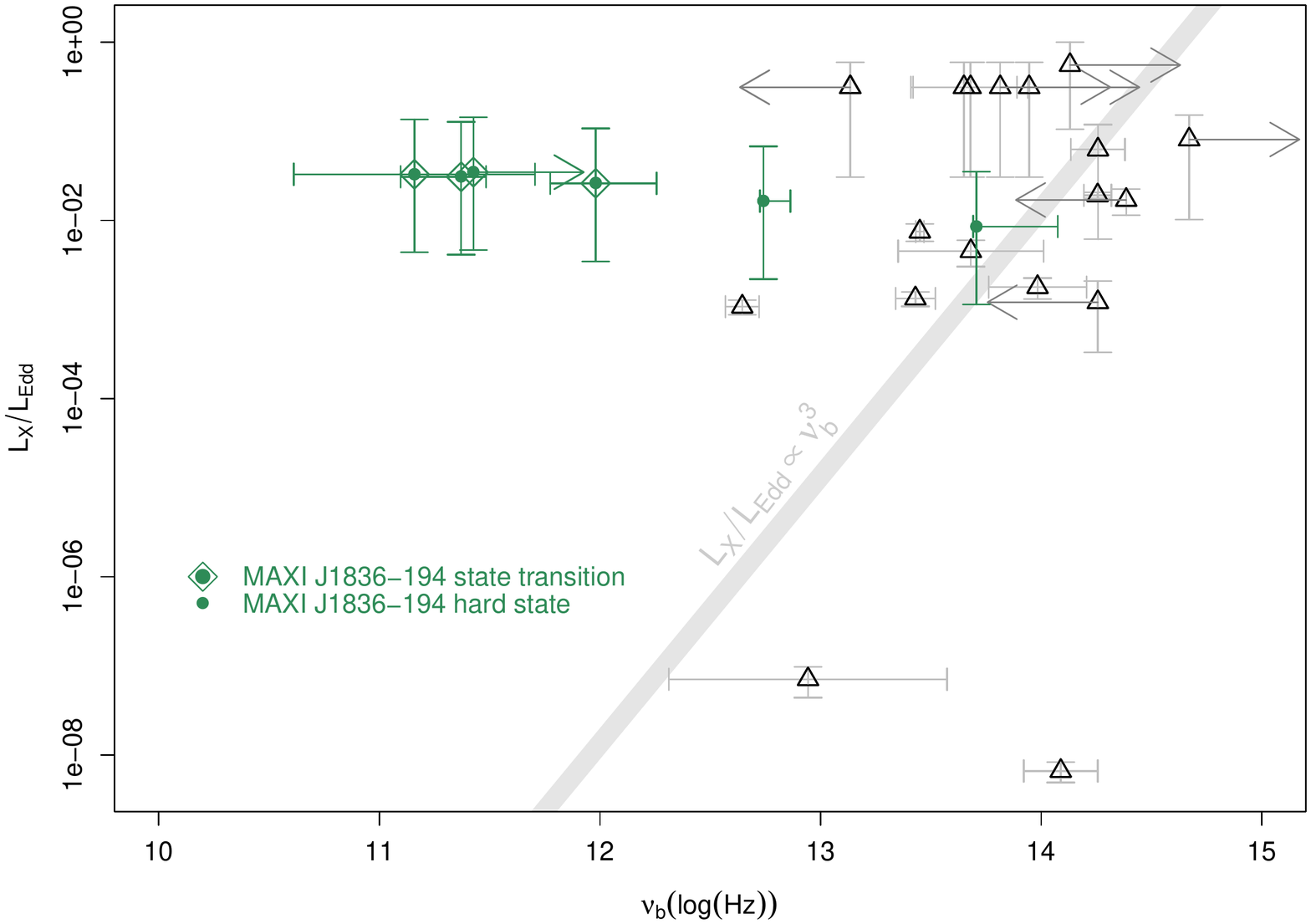}
\caption{Inferred values of jet spectral break frequencies as a function of Eddington scaled luminosity for a sample of 12 BHXRBs (Russell \etal\ 2013a,b). The evolution of MAXI J1836-194, highlighted in green, is in particularly stark contrast with the simple relation expected by basic synchrotron theory plus accretion scaling arguments. 
Data courtesy of Dave Russell and Tom Russell.}
\label{fig:breaks}      
\end{figure*}

Prior to this work, a {\it shifting} jet spectral break frequency
(that is, within the duration of a single outburst) had been observed
for one BHXRB only; GX339--4 (Gandhi \etal\ 2011; Corbel
\etal\ 2013). Russell \etal\ (2013a) reported on the same behavior for
XTE J1118+480 , whose jet break frequency varied by more than one
order of magnitude while the X-ray luminosity change was negligible.
More recently, high cadence, multi-wavelength monitoring of the 2011
outburst of \maxi\ (Russell \etal\ 2013b), showed even more severe
discrepancies between the data and the expected scaling between
$\nu_b$ and \lx; during the early days of the outburst, $\nu_b$ for
\maxi\ moved almost perpendicular to theoretical line (see Fig \ref{fig:breaks}), strongly indicating a much higher degree of
complexity in the processes that regulate the jet special energy
distribution, particularly during the initial phases of the compact
jet recovery. \\

A second, major finding of this multi-wavelength campaign was the
observational inference, for the first time, of the high energy
synchrotron cooling break: $3.2\times 10^{14}\simlt \nu_{c} \simlt
4.5\times 10^{14}$ Hz. Though the cooling break was not directly
visible in the broad-band spectral energy distribution data (see
figure 7 of Russell \etal\ 2013b for a zoom into the mid-IR to X-ray
spectrum of \maxi\ as measured on 2011 Sept 03, along with the inset
showing the Ag 31 and Sept 01 VLT spectrum), the presence of a high
energy break in the above frequency range is necessary in order to
reconcile the near-IR synchrotron power-law spectrum with the strength
of the Balmer absorption lines seen in the nearly simultaneous (within
a day) VLT optical spectrum\footnote{As discussed by Russell
  \etal\ (2013b), strong Balmer absorption lines (with relative depths
  as high as 20 per cent of the continuum flux) are inconsistent with
  a synchrotron-dominated continuum; their presence demands an
  optically thick disc that, at the time of the observations, was
  contributing to about 50 per cent of the optical emission.}. It is
important to stress that, for X-ray binaries, the cooling break is
expected to shift from from the ultraviolet to the X-ray band (above
10 keV) as \lx\ increases from quiescence up to bright, hard X-ray
states (Pe'$~$er \& Markoff 2012), while the reported value for
\maxi\ falls into the optical band. This has been interpreted by
Russell and collaborators as evidence for the jet ``already evolving
early in the outburst, at the same time as the system is brightening
and the disc begins filling in".

The works discussed above serve to illustrate how high cadence,
simultaneous, multi-wavelength monitoring of BHXBRs in outburst --
particularly when the more traditional radio/X-ray approach is
integrated with sensitive spectroscopy at IR and optical frequencies
-- hold the key to making further progress in this field. \\

A new avenue is also close to opening up for the study of X-ray binary 
outbursts: radio selected sources from wide-field searches.
By the end of the current decade (i.e. 2020) the first phase of the SKA
will have been constructed, alongside/incorporating the MeerKAT and
ASKAP telescopes. For the first time large fractions of the sky 
(of order unity steradians) will be surveyed regularly (daily) and
we should begin to find such events first by their radio emission.
The mid-frequency component of the first phase of the SKA, `SKA$_1$ Mid',
should be able to easily detect most galactic XRBs during the hard state 
rising phase, providing plenty of alert for new outbursts, as well as
tracking sources all the way to quiescence, and monitoring XRBs in
other galaxies (Fender 2004; see {\bf www.skatelescope.org} for more
information on the SKA project).

\subsection{Baryons?}

One of the most important uncertainties in the study of relativistic jets ever since
their discovery has been the question of their composition. In nearly all cases the
radiation from jets is synchrotron emission, which only requires the presence of
leptons, leaving us little the wiser as to whether the jets are primarily electron:proton or
electron:positron in composition. The only exception to this case is SS433, which is itself
an highly unusual system which is hard to fit into existing classification schemes, and
in which the jet may become baryon-loaded after launch by interactions with a massive and
dense environment (see Begelman, King \& Pringle 2006 and references therein).

In the past year, Diaz-Trigo et al. (2013) have reported the first evidence for strong
atomic emission lines associated with relativistic jets in a (low-mass) black hole X-ray binary,
4U 1630-47. They report evidence for Doppler-shifted emission lines originating in gas
moving at $\sim 0.7c$, at the same time as optically thin radio emission is observed.
This is a very exciting discovery, although it seems unlikely this can be ubiquitous
for all strong-jet states and it may be only associated with certain rare phases of
the outburst cycle (Neilsen et al. 2014). Regardless of the duration of the active baryon-loaded
jet phase, its presence gives us very important new clues into the power of the jets 
and their role in the mass flow close to the black hole.

\section{Summary}

In this brief review we've attempted to provide a picture of the current empirical
understanding of the coupling between accretion and outflow in stellar mass black holes.
A key development in the past few years have been a clearer picture of the role of
baryons in outflows, both in the systematic coupling of winds to accretion states 
reported in Ponti et al. (2012), and the recent, very exciting, discovery of relativistically
moving baryons in a low-mass X-ray binary (Diaz-Trigo et al. 2013). Understanding these
phenomena more deeply should allow us to really being to understand the flow of mass
and energy near accreting black holes in different accretion states. The investigation
of whether or not black hole spin really powers relativistic jets (part or all of the time)
continues, and the results remain tantalising but unclear (to us). Without doubt this
is one of the most important avenues of research in high-energy astrophysics research right now.
What drives the `radio loud' and `radio quiet' branches in the hard X-ray states
is a mystery (although its not spin!), and whether or not everything (anything?) we learn 
from black hole X-ray binaries is really applicable to AGN remains to be proven.
Much, much, more remains to be done, observationally and theoretically, and this 
will be in part driven by new facilties such as the SKA, and by renewed interests
in the origin of X-ray binary states and accretion cycles.

\begin{acknowledgements}
RPF would like to acknowledge an almost uncountable number of conversations with many collaborators over the past two decades.
During this particular writing period he has learnt from Teo Munoz-Darias, Gabriele Ponti, Mickael Coriat, Dan Plant and Joey Neilsen. EG would like to thank Stephane Corbel for sharing his radio and X-ray luminosity data set, Dave and Tom Russell for sharing their data on spectral break frequencies. 

\end{acknowledgements}


\begin{thebibliography}{}
%
\bibitem{}Begelman M. C., King A. R., Pringle J. E., 2006, MNRAS, 370, 399
\bibitem{}Begelman M. C. \& Armitage P. J., 2014, MNRAS, 782, L18
\bibitem{}Belloni T. M., Motta S. E., Munoz-Darias T., BASI, 2011, 39, 409 
\bibitem{}Blandford R. D., Znajek R. L., 1977, MNRAS, 179, 433
\bibitem{}Blandford R. D., Payne D. G., 1982, MNRAS, 199, 883
\bibitem{}Bradley C. K. \etal, 2007, ApJ, 667, 427
\bibitem{}Brocksopp C. \etal, 2010, MNRAS, 404, 908
\bibitem{}Casella P. G. \etal, 2010, MNRAS, 404, L21
\bibitem{}Calvelo D. E. \etal, 2010, MNRAS, 409, 839
\bibitem{}Chaty S. \etal, 2003, MNRAS, 346, 689
\bibitem{}Corbel S. \etal, 2003, A\&A, 400, 1007
\bibitem{}Corbel S., K\"ording E., Kaaret P., 2008, MNRAS, 389, 1697
\bibitem{}Corbel S. et al., 2013, MNRAS, 431, L107
\bibitem{}Coriat M. \etal, 2011, MNRAS, 414, 677
\bibitem{}Coriat M. , Fender R. P. \& Dubus G., 2012, MNRAS, 424, 1991
\bibitem{}Done C., Gierlinski M., Kubota A., 2007, A\&ARv, 15, 1
\bibitem{}Dunn R. J. H. \etal, 2010, MNRAS, 403, 61
\bibitem{}Falcke H., K\"ording E. \& Markoff S., 2004, A\&A, 414, 895
\bibitem{}Fender R., 2004, New Astronomy Reviews, 48, 1399
\bibitem{}Fender R. P., Gallo E. \& Jonker P. G., 2003, MNRAS, 343, L99 
\bibitem{}Fender R. P., Belloni T. \& Gallo E., 2004, MNRAS, 355, 1105 
\bibitem{}Fender R. P., Maccarone T.  J. \& van Kesteren Z., 2005, MNRAS, 360, 1085
\bibitem{}Fender R. P., 2006,  Jets from X-Ray Binaries. Cambridge Univ. Press, Cambridge, p. 381
\bibitem{}Fender R. P., Gallo E., Russell D. M., 2010, MNRAS, 406, 1425 
\bibitem{}Fender R. P. \& Belloni T., 2012, Science, 337, 540
\bibitem{}Fabian A. C. \etal, 2012, MNRAS, 424, 217
\bibitem{}Gallo E., Fender R. P. \& Pooley G. G., 2003, MNRAS, 344, 60
\bibitem{}Gallo E. \etal, 2005, Nature, 436, 819
 \bibitem{}Gallo E., Fender R. P. \& Hynes R., 2005, MNRAS, 356, 1017
\bibitem{}Gallo E. \etal, 2006, MNRAS, 370, 1351 
\bibitem{}Gallo E. \etal, 2007, ApJ, 670, 600
\bibitem{}Gallo E., Miller B. P. \& Fender R. P., 2012, MNRAS, 423, 590
\bibitem{}Gandhi P. et al., 2011, ApJ, 740, L13
\bibitem{}Heinz S., 2002, A\&A, 388, L40
\bibitem{}Heinz S. \& Sunyaev R., 2003, MNRAS, 343, L59
\bibitem{}Heinz S. \& Grimm H., 2005, ApJ, 633, 384
\bibitem{}Heinz S., Merloni A. \& Schwab J., 2007, ApJ, 658, L9
\bibitem{}Hjellming, R. M. \& Rupen M., 1995, Nature, 6531, 464
\bibitem{}Hjellming, R. M., \& Han, X., 1995, Radio Properties of X-ray binaries, p. 308
\bibitem{}Hynes R. \etal, 2004, ApJ,  611, L125
\bibitem{}Hynes R. \etal, 2006, ApJ,  651, 401
\bibitem{}Hynes R. \etal, 2009, MNRAS, 399, 2239
\bibitem{}Khandai N. \etal, 2014, MNRAS submitted (arXiv:1402.0888)
\bibitem{}K\"ording E. G., Fender R. P. \& Migliari S., 2006, MNRAS, 369, 1451
\bibitem{}K\"ording E. G., Jester S. \& Fender R. P., 2006, MNRAS, 372, 1366
\bibitem{}K\"ording E. G. \etal, 2007, MNRAS, 380, 301
\bibitem{}K\"ording E. G., Jester S. \& Fender R. P., 2008a, MNRAS, 383, 277
\bibitem{}K\"ording E.G. \etal\, 2008b Science, 320, 1318
\bibitem{}King A. L. \etal, 2013a, ApJ, 771, 84, 12
\bibitem{}King A. L. \etal, 2013b, ApJ, 762, 18 
\bibitem{}King A. L. \etal, 2014, ApJ, 784, L2, 6
\bibitem{}Kormendy J. \& Ho, L. 2013, ARAA, 51, 511
\bibitem{}Jamil, O.; Fender, R. P.; Kaiser, C. R., 2010, MNRAS, 401, 394
\bibitem{}Joinet, A.; Kalemci, E.; Senziani, F., 2008, ApJ, 679, 655
\bibitem{}Jonker P. G. et al., 2010, MNRAS, 401, 1255 
\bibitem{}Liu B. F. \etal, 2007, ApJ, 671, 695
\bibitem{}Malzac J., 2013, MNRAS, 429, L20
\bibitem{}Malzac J., Merloni A. \& Fabian A. C., 2004,  MNRAS, 351, 253
\bibitem{}Markoff S., Falcke H. \& Fender R. P., 2001, A\&A, 372, L25
\bibitem{}Markoff S., \etal, 2003, A\&A, 397, 645
\bibitem{}Markoff S., Nowak M. \& Wilms J., 2005, ApJ, 635, 1203
\bibitem{}Markoff S. \&  Nowak M. , 2007, ApJ, 609, 972
\bibitem{}Mart{\'{\i}} J., 2005, MmSAI, 76, 592
\bibitem{}Miller, J. M., Homan, J., Miniutti, G., 2006, ApJ, 652, L113
\bibitem{}Miller, J. M., 2007, ARAA, 45, 441
\bibitem{}McClintock J. E. \& Remillard R. A., 2006, Black Hole Binaries. Cambridge Univ. Press, Cambridge , p. 157
\bibitem{}McClintock J. E., Narayan R., \& Steiner, J. 2013, Space Science Reviews (arXiv:1303.1583)
\bibitem{}McHardy I. M. \etal, 2006, Nature, 444, 730
\bibitem{}Merloni A., Heinz S. \& Di Matteo T., 2003, MNRAS, 345, 1057
\bibitem{}Meyer F., Liu B. F. \& Meyer-Hofmeister E., 2007, A\&A, 463, 1
\bibitem{}Meyer-Hofmeister E. \& Meyer F., 2014, A\&A, 562, 142
\bibitem{}Middleton, Miller-Jones J. C. A. \& Fender R. P., 2014, MNRAS, 439, 1740
\bibitem{}Miller J. M., 2006, Nature,  441, 953
\bibitem{}Miller J. M., 2007, ARAA, 45, 441
\bibitem{}Miller J. M., 2008, ApJ, 680, 1359
\bibitem{}Miller J. M. \etal, 2009, ApJ, 697, 900
\bibitem{}Miller J. M. \etal, 2012, ApJ, 759, L6
\bibitem{}Miller-Jones J. C. A. \etal, 2011, ApJ, 739, L18
\bibitem{}Miller-Jones J. C. A. \etal, 2012, MNRAS, 421, 468
\bibitem{}Mirabel I. F. \& Rodr{\'{\i}}guez L. F., 1994, Nature, 371, 46
\bibitem{}Mirabel I. F. \etal, 1998, A\&A, 330, L9
\bibitem{}Mirabel I. F. \& Rodr{\'{\i}}iguez L. F., 1999, ARAA, 37, 409
\bibitem{}Motta, S. E.; Belloni, T.; Homan, J., 2009, MNRAS, 400, 1603
\bibitem{}Motta S. E. \etal, 2014a, MNRAS, 437, 2554
\bibitem{}Motta S. E. \etal, 2014b, MNRAS, 439, L65
\bibitem{}Munoz-Darias T., Motta S., Belloni T., 2011, MNRAS, 410, 679
\bibitem{}Narayan R., McClintock J. E., 2012, MNRAS, 419, L69
\bibitem{}Neilsen J. \& Lee J.C., 2009, Nature, 458, 481
\bibitem{}Neilsen J. \etal\ 2014, ApJL, 784, L5
\bibitem{}Petrucci P.O. \etal\ 2008, MNRAS, 385, L88
\bibitem{}Pe'$~$er A. \& Markoff S., 2012, ApJ, 753, 177
\bibitem{}Plant D.S.\etal\, 2014, MNRAS, submitted
\bibitem{}Ponti G. \etal, 2012, MNRAS, 422, 11
\bibitem{}Remillard R. \& McClintock J.E., 2006, ARA\&A, 44, 49
\bibitem{}Russell D. M. \etal\ 2003, MNRAS, 371, 13
\bibitem{}Russell D. M. \etal\ 2007, MNRAS,  379, 1401
\bibitem{}Russell D. M. \etal\ 2011a, ApJ, 739, L19
\bibitem{}Russell D. M. \etal\ 2011b, MNRAS, 416, 2311
\bibitem{}Russell D. M., Gallo E. \& Fender R., 2013, MNRAS, 431, 405 
\bibitem{}Russell D. M. et al., 2013a, ApJ, 768, L35 
\bibitem{}Russell D.M., Shahbaz T., 2014, MNRAS, 438, 2083
\bibitem{}Russell T. D. \etal\ 2013b, MNRAS in press (2014MNRAS.tmp..180R)
\bibitem{}Reis R. C., Fabian A. C. \& Miller J. M., 2010, MNRAS, 402, 836
\bibitem{}Reynolds C. S., 2013, Space Science Reviews (arXiv:1302.3260)
\bibitem{}Reynolds M. T. \& Miller J. M., 2013, ApJ, 769, 16
\bibitem{}Soleri P., Belloni T., CAsella P., 2008, MNRAS, 383, 1089
\bibitem{}Soleri P. \& Fender R., 2011, MNRAS, 413, 2268
\bibitem{}Steiner J. F., McClintock J. E. \& Narayan R., 2013, ApJ, 762, 104
\bibitem{}Stella L. \& Vietri M., 1999, Phys. Rev. Lett., 82, 17
\bibitem{}Tchekhovskoy A. \etal, 2010, ApJ, 711, 50
\bibitem{}van der Laan H., 1966, Nature, 211, 1131
\end{thebibliography}
\end{document}